\documentclass[]{article}
\usepackage{amsmath}
\usepackage{amssymb}
\usepackage{amsfonts}
\usepackage{graphicx}
\usepackage[margin=1.0in]{geometry}
\usepackage{amsthm}
\theoremstyle{definition}

\newcommand{\captionfonts}{\small}

\newtheorem{definition}{Definition}

\makeatletter  
\long\def\@makecaption#1#2{%
  \vskip\abovecaptionskip
  \sbox\@tempboxa{{\captionfonts #1: #2}}%
  \ifdim \wd\@tempboxa >\hsize
    {\captionfonts #1: #2\par}
  \else
    \hbox to\hsize{\hfil\box\@tempboxa\hfil}%
  \fi
  \vskip\belowcaptionskip}
\makeatother   

\title{Second-order Price Dynamics: Approach to Equilibrium with Perpetual Arbitrage}
\author{Eric Kemp-Benedict\\
Stockholm Environment Institute\\
eric.kemp-benedict@sei-international.org}

\begin{document}

\maketitle
\bibliographystyle{plain}

\begin{abstract}
The notion that economies should normally be in equilibrium is by now well-established; equally well-established is that economies are almost never precisely in equilibrium. Using a very general formulation, we show that under dynamics that are second-order in time a price system can remain away from equilibrium with permanent and repeating opportunities for arbitrage, even when a damping term drives the system towards equilibrium. We also argue that second-order dynamic equations emerge naturally when there are heterogeneous economic actors, some behaving as active and knowledgeable arbitrageurs, and others using heuristics. The essential mechanism is that active arbitrageurs are able to repeatedly benefit from the suboptimal heuristics that govern most economic behavior.

\vspace{2em}

\end{abstract}

\section{Introduction}
Dynamic macroeconomics proposes that economic actors set prices based on gaps -- perceived or real -- between supply and demand. The aggregate gap between supply and demand is determined by the current vector of prices $p$, which has as many elements as there are commodities, through the aggregate demand function $\xi(p)$, a vector-valued function of the same dimensionality as $p$. In the standard formulation, the prices in one time period are set depending on the excess demand that results from the prices in the previous period. In continuous time, this leads to a first-order dynamic equation,
\begin{equation}\label{eqn_firstorder_canonical}
\frac{dp}{dt} = \kappa\xi(p).
\end{equation}
When excess demand for a product or commodity is positive, sellers or speculators raise the price, judging that they could gain extra income. When excess demand is negative, sellers or speculators lower the price, hoping to stimulate demand sufficiently to raise their total income. At equilibrium, excess demand is zero and prices stop changing. Importantly, at equilibrium there is no opportunity for arbitrage -- selling one commodity or product at what is judged to be an elevated price, or buying an under-priced commodity or product, in order to gain a temporary advantage. Moreover, opportunities for arbitrage should rapidly disappear as active arbitrageurs take advantage of them \cite{fama_behavior_1965}.

The historical development of the dynamic theory of prices is generally thought to be unsatisfactory because, after some initially promising results, it could not guarantee stability \cite{jordan_instability_1986,saari_iterative_1985,scarf_examples_1960}. Part of the argument has a solid and unavoidable foundation: the so-called SMD theorem elaborated in separate papers by Sonnenschein \cite{sonnenschein_market_1972}, Mantel \cite{mantel_characterization_1974}, and Debreu \cite{debreu_excess_1974}. These papers collectively showed that there are almost no constraints on the excess demand function, no matter how regular the behavior of individual economic actors. The other part of the instability argument flows from the first-order dynamics.

Macroeconomics today is dominated by the theory of General Equilibrium (e.g., \cite{ginsburgh_structure_2002}). General Equilibrium theory seeks to avoid the problems of dynamic macroeconomics by, essentially, not providing any dynamics at all. Instead, exogenous changes in supply and demand functions, combined with exogenous ``shocks'' push the economy away from a preexisting equilibrium. Then, following an unknown process, a new equilibrium is formed quickly enough that it can be treated as instantaneous from the point of view of the theory. This can be interpreted as a modest and sensible approach to deal with an unsatisfactory dynamic theory. However, despite the popularity and appeal of General Equilibrium theory, it inherits the shaky foundation of dynamic economics, because the SMD theorem appears to make any price adjustment mechanism -- even an unspecified one -- problematic \cite{ackerman_still_2001}. Moreover, as Fisher \cite{fisher_disequilibrium_1989} points out, it is difficult to analyze equilibria without a theory of disequilibrium.

In this paper, and building on a previous paper \cite{kemp-benedict_second-order_2011}, we challenge the idea that price dynamics should be first-order in time. Following Samuelson's \cite{samuelson_foundations_1947} formulation of Walras' \textit{t\^atonnement} adjustment process \cite{walras_elements_1954}, economists have assumed an adjustment process that is first-order in time, as in Equation \ref{eqn_firstorder_canonical}. (See, e.g., \cite{flaschel_dynamic_1997,wickens_macroeconomic_2012}).\footnote{Fisher \cite{fisher_disequilibrium_1989} provides a theory of prices without an explicit time dependence. He shows that stability is guaranteed very generally in theories that feature No Favorable Surprise; that is, opportunities for arbitrage, real or perceived, do not continue to appear. The theory presented in this paper is not of this class. Instead, prices approach equilibrium but never reach it because some actors use sub-optimal heuristics that lead to recurring Favorable Surprises.} In this paper we argue that it is difficult to maintain first-order price dynamics and that, from different starting points, it is easy to reach an equation that is second-order in time. We then show that solutions to the resulting dynamic equations have interesting properties that are quite different from those of a first-order dynamic equation. Near an equilibrium point, the dynamic equations are like those governing a collection of damped harmonic oscillators. While economic systems have, in the past, been modeled as coupled harmonic oscillators (see, e.g., \cite{ikeda_coupled_2011}), it is unusual. This paper argues that models with second-order time dynamics arise so naturally that they are a much better starting point than first-order equations when studying how economic actors determine prices.

Before proceeding, we acknowledge that there are compelling arguments against looking for equilibria in economics at all -- at least equilibria as Walras described them. In a lecture subsequently published in \textit{The Economic Journal}, Kaldor \cite{kaldor_irrelevance_1972} argued that theories of equilibrium prices described economies where creative process were strangely erased. He noted in particular that they could not account for the obvious fact of economies of scale in industrial production. In contemporary terms, the economy that Kaldor described can be seen as a complex adaptive system, in which creative forces alter demands for materials in a dynamic and reciprocal way and actors absorb fluctuations in the system by holding or releasing stocks, while any (non-unique and fleeting) stable structure emerges from internal dynamics. A quite distinct critique was offered by Keynes \cite{keynes_general_2010}, who sought to explain economic depressions through a ``general'' economic theory, in contrast to the ``special'' theory that explained economic activity in normal times. His theory explains how the economy can remain in a stable state -- an equilibrium -- with a large gap between the supply and demand for labor and large unsold stocks of goods.

This paper does not fundamentally challenge theories of (Walrasian) price equilibria, as did Kaldor and Keynes. Instead, it offers a ``special'' theory, in the sense of Keynes, and over short enough times that the creative processes noted by Kaldor are unimportant when compared to price-setting behavior. The surprise is that this special, limited theory, which differs only modestly from prior theories of price-setting behavior, results in price dynamics that approach, but rarely reach, equilibrium and that offers, as a consequence of its dynamics, endless opportunities for arbitrage.

\section{Routes to Second-order Price Dynamics}
In this section we argue that second-order price dynamics emerge naturally when there are heterogeneous actors, some of whom act immediately using standard first-order price setting and others of whom act in a subsequent period. In two of our examples we assume that one group -- the arbitrageurs -- informs itself about the market and actively seeks opportunities to take advantage of positive or negative excess demand, while a second group -- the majority of economic actors -- uses heuristics to guide its actions. The third example considers arbitrageurs operating under uncertainty at different times.

The use of heuristics in economic reasoning, and in particular for setting prices, is well established. Keynes postulated that employers found it difficult to lower wages in nominal terms, a phenomenon he called a ``sticky wage'' \cite{keynes_general_2010}. Later work focused more on sticky prices than wages. These theories had little empirical backing until Blinder \cite{blinder_sticky_1994,blinder_asking_1998} carried out a survey of firms and directly asked managers about their pricing behavior. He and his colleagues found that prices are indeed sticky, with a median change of 1.3 times per year; they also found that firms set prices based on heuristics such as waiting for other firms to change their prices first and basing prices on costs. The importance of heuristics in economics, and more generally of the field of Behavioral Economics (see, e.g., \cite{camerer_behavioral_2004}), was formally recognized when the 2002 Nobel Prize in Economics was awarded to Kahneman for his work with Tversky \cite{tversky_judgment_1974,kahneman_maps_2002}. Even in financial markets, which are widely believed to efficiently and rapidly drive prices toward equilibrium, Bouchard et al. \cite{hens_how_2009} found evidence that actual price setting is slow and noisy, characterized by the gradual absorption of imperfect information, while in an important sequence of papers Gabaix and his colleagues unified stylized facts about the distribution of returns, trading volume, price impact, and large investors using a model in which different-sized investors exhibit different trading behavior \cite{gabaix_theory_2003,gabaix_institutional_2006,gabaix_unified_2007}.

One heuristic, which was observed in an artificial experimental setting by Ostrom and her colleagues \cite{ostrom_rules_1994} and in the behavior of firms in the survey by Blinder, is that some actors, group $a$, set their prices immediately, while others, group $b$, wait and set their price in the next time period based on the change in average prices from the previous time period. Following \cite{kemp-benedict_second-order_2011}, and setting $\kappa = 1$ in Equation \ref{eqn_firstorder_canonical} for convenience, we write
\begin{subequations}
\begin{align}
\Delta p_{a,t} &= \mu \xi(\bar{p}_{t-1}),\\
\Delta p_{b,t} &= \nu\Delta{\bar{p}}_{t-1}.
\end{align}
\end{subequations}
If a fraction $f_a$ of sellers are of type $a$, and a fraction $f_b = 1-f_a$ are of type $b$, and we compute $\Delta\bar{p}_t = f_a \Delta p_{a,t} + f_b \Delta p_{b,t}$, then after substituting we get
\begin{equation}\label{eqn_barp_heur1_init}
\Delta\bar{p}_t = f_a\mu \xi(\bar{p}_{t-1}) + f_b\nu\Delta{\bar {p}}_{t-1}.
\end{equation}
Next, writing $\Delta\bar{p}_{t-1}$ as
\begin{equation}
\Delta\bar{p}_{t-1} = \Delta\bar{p}_t - \Delta^2\bar{p}_t,
\end{equation}
where $\Delta^2{p}_t\equiv \Delta\bar{p}_t - \Delta\bar{p}_{t-1}$ is a second-order difference, Equation \ref{eqn_barp_heur1_init} can be rearranged to give
\begin{equation}\label{eqn_barp_heur1_final}
f_b\nu\Delta^2\bar{p}_t = f_a\mu \xi(\bar{p}_{t-1}) - (1 - f_b\nu)\Delta{\bar {p}}_t.
\end{equation}
This equation, which has a difference of differences on the left-hand side, is second-order in time; if $f_b$ or $\nu$ goes to zero, then it returns to a first-order equation. Also, if $f_b\nu < 1$ then the second term on the right-hand side is a damping term, which dissipates economic activity and drives the system toward equilibrium.

A second type of heuristic is the collection of trend-spotting methods used in technical trading. While in theory these techniques should not work (see, e.g., \cite{fama_behavior_1965}), they are applied in practice, and are sufficiently successful to support a large and growing number of books on the topic (e.g.,\cite{chen_essentials_2010,pring_technical_2002}). Since practitioners may still be deluded, more convincing is the evidence provided by Garzarelli et al. \cite{garzarelli_memory_2011}, who, in contrast to theoretical expectations, observed signals of technical trading in financial data. Technical traders expect prices of traded assets to go through recurring periods of accelerating, and then slowing, rises and falls. An accelerating rise in price, or a slowing fall, is driven by the activities of ``bulls'' who are optimistic about future price trends; a slowing rise in price, or an accelerating fall, is driven by the activities of ``bears''. The technical trader tries to anticipate where the bulls and bears are going, and follow along. Using the same notation as above, suppose that the bulls comprise group $a$ and the bears group $b$. In this case we could write
\begin{subequations}
\begin{align}
\Delta p_{a,t} &= \lambda \xi(\bar{p}_{t-1}) + \nu\Delta{\bar{p}}_{t-1},\\
\Delta p_{b,t} &= \mu \xi(\bar{p}_{t-1}) - \gamma\Delta{\bar{p}}_{t-1}.
\end{align}
\end{subequations}
With these behaviors, both bulls and bears respond to perceived opportunities, as reflected in excess demand, but bulls are encouraged by a rising market, whereas bears are discouraged by it and ready to judge a rapidly-growing market as overheated. Calculating the average change in price as before, these behaviors result in the following equation,
\begin{equation}
\Delta\bar{p}_t = (f_a\lambda + f_b\nu) \xi(\bar{p}_{t-1}) + (f_a\nu - f_b\gamma)\Delta{\bar {p}}_{t-1}.
\end{equation}
This equation has the same structure as Equation \ref{eqn_barp_heur1_init}, and like it can be rearranged to give a second-order equation similar to Equation \ref{eqn_barp_heur1_final}.

A third possibility is that everyone is an arbitrageur, but different arbitraguers respond to price signals at different times (as in \cite{abreu_bubbles_2003}). In the case that group $a$ responds in the next time period, while group $b$ delays a further time period before responding, the change in price in a given time period is
\begin{equation}
\Delta p_t = f_a \xi(\bar{p}_{t-1}) + f_b \xi(\bar{p}_{t-2}).
\end{equation}
Expanding $\xi(\bar{p}_{t-2})$ about the point $p_{t-1}$ to first order in a Taylor series, this becomes
\begin{equation}
\Delta p_t \approx \xi(\bar{p}_{t-1}) - f_b \Delta p_{t-1}\cdot\nabla\xi(\bar{p}_{t-1}),
\end{equation}
where $\nabla\xi(p)$ is the vector derivative of $\xi(p)$ with respect to price. As with the other examples, this can be expressed as a second order equation, possibly with damping, in which the coefficient on the damping term is a matrix that depends on the price vector.

\section{Second-order Dynamics in Continuous Time}
We consider dynamics in continuous time and in which prices can take on any real value. Before proceeding we note that both of these assumptions are problematic. It is well known that discrete-time dynamics can be different from continuous-time dynamics (see, e.g., \cite{saari_iterative_1985}), and arguably discrete-time dynamics are more relevant for economic analysis. The discrete-time behavior of second-order price dynamics is discussed in \cite{kemp-benedict_second-order_2011}; here we focus on the solutions to the continuous-time dynamical theory, which have interesting properties that are worth exploring. The assumption that prices can take on any real value within a compact space appears to be generally accepted as a reasonable approximation to reality. However, Nadal \cite{nadal_behind_2004} points out that prices cannot take on any real value and they in fact lie within a non-compact space. This is an interesting point deserving more attention, but in this paper we make the conventional assumption that prices can be represented as a real-valued vector.

The continuous-time equation we shall consider is
\begin{equation}\label{eqn_secorder_basic}
\frac{d^2p}{dt} + p\left| \frac{dp}{dt}\right|^2 = \kappa\xi(p) - \gamma\cdot\frac{dp}{dt},
\end{equation}
where, as explained in \cite{kemp-benedict_second-order_2011}, the second term on the left-hand side of the equation ensures that the price vector stays on the hypersphere defined by $|p|^2\equiv\sum_{i=1}^n p_i^2 = 1$.\footnote{To understand the additional term in Equation \ref{eqn_secorder_basic}, note that $p\cdot(\ddot{p} + p|\dot{p}|^2) = p\cdot\ddot{p} + |p|^2|\dot{p}|^2 = (1/2)d|p|^2/dt^2 - |\dot{p}|^2 + |p|^2|\dot{p}|^2 = 0$, where an overdot indicates a time derivative. The final result is zero because $|p|^2 = 1$. } This price normalization is required because the overall magnitude of the price vector carries no information, and only relative prices are important. In this equation we take $\kappa$ to be a scalar and $\gamma$ to be a diagonal matrix with all positive elements -- that is, damping may vary from one commodity to another, but while the price of one commodity may affect the price of another through the excess demand function, the rate of change of the price of that commodity only affects subsequent changes in its own price through damping. This assumption is not essential, but it simplifies some subsequent calculations.

The excess demand function for an $n$-commodity economy is a continuous function that takes a vector of prices, $p$, with $n$ positive elements and produces a vector with $n$ elements. Mathematically, this is expressed as $\xi: (\mathbb{R}^+)^n \rightarrow \mathbb{R}^n$. More specifically, as noted above, because the magnitude of the price vector carries no information the price vector can be restricted to the positive portion of the hypersphere $S^+_n$ defined by $|p|^2 = 1$. Finally, Walras' Law states that the excess demand function is orthogonal to the price vector, so that $p\cdot\xi(p)=0$ for all combinations of prices, so that, in monetary terms, the economy is always in overall equilibrium, even though individual commodities might be out of equilibrium.\footnote{This is not true in Keynesian economics. However, as stated earlier, this paper addresses what Keynes called the ``special'' theory of economics, where Walras' Law is a reasonable assumption.} Combining these observations we recover the definition of an excess demand function offered by Debreu \cite{debreu_excess_1974}. In the terms defined here, the definition can be given as
\begin{definition}[Excess demand function]
A continuous function $\xi: S^+_n \rightarrow \mathbb{R}^n$ such that $p\cdot\xi(p) = 0$ for all $p\in S^+_n$.
\end{definition}
The essence of the SMD theorem is that almost any such function is compatible with some combination of well-behaved individual demand functions. So, while individual demand functions may have nice properties such as marginally declining demand as prices rise, consistent preference rankings, and so on, the aggregate demand function can have almost any behavior at all as long as it is a continuous vector-valued function on the positive portion of the price hypersphere.

Fortunately, the properties of being a continuous vector-valued function on a compact space (\textit{pace} Nadal) are all that is required for the Helmholtz-Hodge decomposition theorem to apply. According to this theorem, the excess demand function can be written as a sum of the gradient of a scalar potential and another, divergence-free, vector field. That is,
\begin{equation}
\xi(p) = -\nabla\phi + A,
\end{equation}
where $\nabla\cdot A = 0$. Substituting this equation into Equation \ref{eqn_secorder_basic}, we have
\begin{equation}
\frac{d^2p}{dt} + p\left| \frac{dp}{dt}\right|^2 = -\kappa\nabla\phi(p) + \kappa A(p) - \gamma\cdot\frac{dp}{dt}.
\end{equation}
From this equation we can get an interesting result by dot-multiplying on the left by the rate of change of the price vector $dp/dt$. This gives
\begin{equation}\label{eqn_secorder_leftmult_init}
\frac{dp}{dt}\cdot\frac{d^2p}{dt} + \frac{dp}{dt}\cdot p\left| \frac{dp}{dt}\right|^2 = -\kappa\frac{dp}{dt}\cdot\nabla\phi(p) + \frac{dp}{dt}\cdot\kappa A(p) - \frac{dp}{dt}\cdot\gamma\cdot\frac{dp}{dt}.
\end{equation}
Taking the first three of these terms, we find that
\begin{subequations}
\begin{align}
\frac{dp}{dt}\cdot\frac{d^2p}{dt} &= \frac{1}{2}\frac{d}{dt}\left| \frac{dp}{dt}\right|^2\label{subeqn_kineticterm}\\
\frac{dp}{dt}\cdot p\left| \frac{dp}{dt}\right|^2 &= \frac{1}{2}\left| \frac{dp}{dt}\right|^2\frac{d\left| p\right|^2}{dt}\label{subeqn_vanishing_orth}\\
\frac{dp}{dt}\cdot\nabla\phi(p) &= \frac{d}{dt}\phi(p)\label{subeqn_dpotdt}
\end{align}
\end{subequations}

The second of these equations, Equation \ref{subeqn_vanishing_orth}, is zero because the price vector is constrained to remain on the surface $|p|^2=1$. The third, Equation \ref{subeqn_dpotdt}, is true as long as the potential is not changing in time.\footnote{This is consistent with the short times we are concerned with in this paper, but also shows how the theory can be generalized to include a changing economic environment, by adding explicit time dependence to the potential function.} Substituting Equations \ref{subeqn_kineticterm} and \ref{subeqn_dpotdt} into Equation \ref{eqn_secorder_leftmult_init} and rearranging gives
\begin{equation}\label{eqn_secorder_totenergy}
\frac{d}{dt}\left(\frac{1}{2}\left| \frac{dp}{dt}\right|^2 + \kappa\phi(p)\right)=\frac{dp}{dt}\cdot\kappa A(p) - \frac{dp}{dt}\cdot\gamma\cdot\frac{dp}{dt}.
\end{equation}

This term in parentheses on the left-hand side of Equation \ref{eqn_secorder_totenergy} has a precise physical analogue: the total energy of a body in a conservative potential. If the divergence-free portion of the excess demand function, $A(p)$ were equal to zero, then the equation would become
\begin{equation}
\frac{d}{dt}\left(\frac{1}{2}\left| \frac{dp}{dt}\right|^2 + \kappa\phi(p)\right)= -\frac{dp}{dt}\cdot\gamma\cdot\frac{dp}{dt}, \text{   for }A=0.
\end{equation}
In this special case it is seen that, just as in the physical analogue, the damping term gradually dissipates the total energy. This will eventually bring the system to rest at a local minimum of the potential. That is, the system will approach equilibrium. However, it is easy to see that in general $A(p)$ will not be zero. If it were, then the Jacobian of the excess demand function, $\partial_i\xi_j(p)$ would be equal to $-\partial_i\partial_j\phi(p)$ and, since partial derivatives commute, the Jacobian would be symmetric. A symmetric Jacobian for the demand function would mean that the influence of the price of commodity $i$ on demand for commodity $j$ would be equal to the influence of the price of commodity $j$ on commodity $i$, and there is no reason to expect such an outcome for most pairs of commodities.\footnote{Note that this does not imply that the Jacobian of $A(p)$ must be \textit{anti}symmetric. The Helmholtz-Hodge theorem only says that the trace of the Jacobian of $A(p)$ vanishes.}

We are therefore left with the full expression in Equation \ref{eqn_secorder_totenergy}. This is more difficult to analyze than the case of a conservative potential, but also more interesting. To make some progress we suppose that there is a price trajectory $p^*(t)$ that travels in a loop that repeats at some point (possibly after wandering for quite a while), such that $p^*(T)=p^*(0)$ and $dp^*(T)/dt = dp^*(0)/dt$ at some time $T$. In this case we can integrate across the time it takes to traverse the loop to find
\begin{equation}
\left.\left(\frac{1}{2}\left| \frac{dp}{dt}\right|^2 + \kappa\phi(p)\right)\right|^T_0 =\kappa \oint_{p^*} dp \cdot A(p) - \oint_{p^*} dp\cdot\gamma\cdot\frac{dp}{dt}.
\end{equation}
Because the loop repeats itself, the left-hand side vanishes: we have traveled around a loop in a conservative potential and ended up where we started, and the total energy is the same in the end as it was at the start. We therefore find
\begin{equation}\label{eqn_secorder_lineintegral}
\kappa \oint_{p^*} dp \cdot A(p) = \oint_{p^*} dp\cdot\gamma\cdot\frac{dp}{dt}.
\end{equation}
The economically interesting aspect of this equation is that if the divergence-free portion of the excess demand function is not zero, then over a continuously repeating loop of price changes, the ``energy'' gained from those price changes can counteract the dissipative effect of the damping term. In human terms what this means is that the efforts of some economic actors to stop the spinning wheel of price changes (the damping term on the right) merely reopen opportunities for arbitrage that active arbitrageurs can take advantage of (the term on the left).

\section{A Two-price System}
To make the theory presented in this paper more concrete we consider a system with two commodities that are normal goods and partial complements. The excess demand function is given by
\begin{subequations}
\begin{align}
\xi_1(p) &= -\alpha (p_1 - \hat{p}_1) + \left(\beta + \delta\right)(p_2 - \hat{p}_2),\\
\xi_1(p) &= -\alpha (p_2 - \hat{p}_2) + \left(\beta - \delta\right)(p_1 - \hat{p}_1),
\end{align}
\end{subequations}
where $\hat{p} = (\hat{p}_1,\hat{p}_2)$ is the equilibrium price. The own-price coefficient $\alpha$ is assumed to be the same for both commodities to simplify later calculations. The parameters $\alpha$ and $\beta$ are assumed to be positive, but $\delta$ can be either sign -- indeed, by simply switching the labels on the commodities $\delta$ can pass from one sign to the other. The symmetric part of the excess demand function can be derived from a potential,
\begin{equation}
\phi(p) = \frac{1}{2}\alpha\left[(p_1 - \hat{p}_1)^2 + (p_2 - \hat{p}_2)^2\right] - \beta (p_1 - \hat{p}_1)(p_2 - \hat{p}_2).
\end{equation}
However, the asymmetric term, proportional to $\delta$, cannot be captured in the potential.

We assume that prices are a small distance away from the equilibrium point and define the deviation vector $q\equiv p - \hat{p}$. We also assume that the dissipative coefficient is the same for both commodities: $\gamma_1 = \gamma_2 \equiv \gamma$. With these assumptions, and to first order in $q$, the second-order dynamic equation, Equation \ref{eqn_secorder_basic}, becomes
\begin{subequations}\label{eqn_secorder_twoprice}
\begin{align}
\frac{d^2q_1}{dt} &= -\kappa\alpha q_1 + \kappa(\beta + \delta)q_2 - \gamma\frac{dq_1}{dt},\\
\frac{d^2q_2}{dt} &= -\kappa\alpha q_2 + \kappa(\beta - \delta)q_1 - \gamma\frac{dq_2}{dt}.
\end{align}
\end{subequations}
These are the equations for a set of coupled harmonic oscillators, which can have quite complex behavior. Because the purpose of this section is to gain insight rather than to simulate an actual market, we consider two cases that have relatively simple behavior: $\delta = 0$ and $\beta = 0$.

\subsection{Conservative dynamics: $\delta = 0$}
Considering first the case in which $\delta=0$, which corresponds to motion in a conservative potential, Equations \ref{eqn_secorder_twoprice} become
\begin{subequations}
\begin{align}
\frac{d^2q_1}{dt} &= -\kappa\alpha q_1 + \kappa\beta q_2 - \gamma\frac{dq_1}{dt},\\
\frac{d^2q_2}{dt} &= -\kappa\alpha q_2 + \kappa\beta q_1 - \gamma\frac{dq_2}{dt}.
\end{align}
\end{subequations}
Defining the linear combinations of the price deviations $y1\equiv q_1 + q_2$ and $y_2\equiv q_1 - q_2$ we find, by taking the sum and differences of these equations,
\begin{subequations}\label{eqn_nodelta_lincomb}
\begin{align}
\frac{d^2y_1}{dt} &= -\kappa(\alpha + \beta) y_1 - \gamma\frac{dy_1}{dt},\\
\frac{d^2y_2}{dt} &= -\kappa(\alpha - \beta) y_2 - \gamma\frac{dy_2}{dt}.
\end{align}
\end{subequations}
These equations are ordinary differential equations and can be solved using standard techniques. The most straightforward is to propose \textit{ans\"atze},
\begin{subequations}
\begin{align}
y_1 = A_1 e^{\omega_1 t},\\
y_2 = A_2 e^{\omega_2 t}.
\end{align}
\end{subequations}
Substituting these into Equations \ref{eqn_nodelta_lincomb} gives a pair of quadratic equations,
\begin{subequations}
\begin{align}
\omega_1^2 &= -\kappa(\alpha + \beta) - \gamma\omega_1,\\
\omega_2^2 &= -\kappa(\alpha - \beta) - \gamma\omega_2,
\end{align}
\end{subequations}
which are solved by
\begin{subequations}
\begin{align}
\omega_1 &= -\frac{\gamma}{2}\left(1 \pm \sqrt{1 - \kappa\frac{\alpha+\beta}{\gamma^2}}\right),\\
\omega_2 &= -\frac{\gamma}{2}\left(1 \pm \sqrt{1 - \kappa\frac{\alpha-\beta}{\gamma^2}}\right).
\end{align}
\end{subequations}
If $\beta < \alpha$ then these solutions always decay; if not, then the equilibrium is not stable, and there is a growing mode. Also, if $\kappa(\alpha + \beta) > \gamma^2$, then $\omega_1$ has an imaginary part, and the price vector spirals around the equilibrium even as it is decaying toward the equilibrium.

The result from this calculation is consistent with the general result given in the previous section -- if the excess demand function can be derived from a potential, then the price decays toward a stable equilibrium.

\subsection{Dynamics with asymmetric price responses: $\beta = 0$}
Next we consider the case in which $\beta = 0$. In this case Equations \ref{eqn_secorder_twoprice} become
\begin{subequations}
\begin{align}
\frac{d^2q_1}{dt} &= -\kappa\alpha q_1 + \kappa\delta q_2 - \gamma\frac{dq_1}{dt},\\
\frac{d^2q_2}{dt} &= -\kappa\alpha q_2 - \kappa\delta q_1 - \gamma\frac{dq_2}{dt}.
\end{align}
\end{subequations}
Rather than solving these equations fully, we manipulate them into a form in which we can recognize another physical analogue. To do this, we multiply the first equation by $q_2$, the second by $q_1$, and then subtract. The result is
\begin{equation}\label{eqn_rot_basic}
q_2\frac{d^2q_1}{dt}-q_1\frac{d^2q_2}{dt} = \kappa\delta\left(q_1^2 + q_2^2\right) - \gamma\left(q_2\frac{dq_1}{dt} - q_1\frac{dq_2}{dt}\right).
\end{equation}
Next, we note that
\begin{subequations}
\begin{align}
\frac{d}{dt}\left(q_2\frac{dq_1}{dt} - q_1\frac{dq_2}{dt}\right) &= q_2\frac{d^2q_1}{dt}-q_1\frac{d^2q_2}{dt} + \frac{dq_2}{dt}\frac{dq_1}{dt} - \frac{dq_1}{dt}\frac{dq_2}{dt}\\
&= q_2\frac{d^2q_1}{dt}-q_1\frac{d^2q_2}{dt}.
\end{align}
\end{subequations}
Therefore, Equation \ref{eqn_rot_basic} becomes
\begin{equation}\label{eqn_rot_angmom_intermediate}
\frac{d}{dt}\left(q_2\frac{dq_1}{dt} - q_1\frac{dq_2}{dt}\right) = -\kappa\delta\left(q_1^2 + q_2^2\right) - \gamma\left(q_2\frac{dq_1}{dt} - q_1\frac{dq_2}{dt}\right).
\end{equation}
Defining the ``angular momentum'' $L$ of the deviation of the price vector around the equilibrium point as
\begin{equation}
L \equiv q_2\frac{dq_1}{dt} - q_1\frac{dq_2}{dt},
\end{equation}
and the magnitude of the price deviation $|q|^2 \equiv q_1^2 + q_2^2$, we can rewrite Equation \ref{eqn_rot_angmom_intermediate} as
\begin{equation}
\frac{dL}{dt} = \kappa\delta|q|^2 - \gamma L.
\end{equation}
Note that if $\delta$ is set to zero in this equation then the angular momentum $L$ simply decays, consistent with the case $\delta=0$ elaborated above and the general remarks in the previous section. However, if $\delta$ is not equal to zero then there is a solution with a nonzero but constant angular momentum, where
\begin{equation}
L = \kappa\frac{\delta}{\gamma}|q|^2.
\end{equation}
This expresses the same phenomenon as in Equation \ref{eqn_secorder_lineintegral}, where prices keep circulating around an equilibrium because of recurring arbitrage opportunities opened up by the second-order time dynamics. The ``angular momentum'' of the price movements decreases as the damping $\gamma$ increases, and increases with greater asymmetries $\delta$ between the price responses of different commodities.

\section{Conclusion}
In this paper we have shown that if there are heterogeneous economic actors, some behaving as active arbitrageurs and others using heuristics, then it is not difficult to generate price dynamics that are second-order in time. More strongly, we conclude that it is difficult to maintain price dynamics that are first-order in time. In general, the variety of possible responses by sellers and speculators in setting prices could lead to multiple time lags of varying length, producing even more complex dynamics.

When prices are governed by second-order dynamics with damping -- where damping reflects heterogeneity of price responses, including the disinclination of some economic actors to continue changing prices -- even the very loose constraints on the excess demand function placed by the SMD theorem leads to interesting results. The Helmholtz-Hodge theorem allows us to decompose the excess demand function in a revealing way, and we find a combination of convergent and non-convergent behavior. The non-convergent behavior emerges from that part of the excess demand function that cannot be expressed as the gradient of a scalar potential; in the two-price system used as an example in this paper, it appeared as an asymmetry in the Jacobian of the excess demand function. In contrast to first-order price dynamics, second-order dynamics with damping result in an endless spiral of price-setting, in which those economic actors who rely on heuristics -- or even partial and fuzzy information -- repeatedly create arbitrage opportunities for active and knowledgeable arbitrageurs.

This interesting result -- of approaching but never quite reaching equilibrium -- is inherent to the second-order dynamics. It means that the system is stable overall, because prices stay near to equilibria, but are never at rest, because of recurring arbitrage opportunities. Rather than an equilibrium point there is a zone of price movements around the equilibrium. Because this behavior is inherent to the dynamics, arbitrage opportunities are not created by noise or shocks, but rather by normal economic activities. There are also, of course, noise and shocks in real markets that could disrupt the dynamics described in this paper. For example, if damping is so large that prices move very slowly around an equilibrium point, then the characteristic recurrence time of shocks may be shorter than the characteristic periodicity of the internal dynamics. In this case the price movements from the internal dynamics would be unobservable, and the system would be dominated by noise. In cases where damping is very low, the internal dynamics could dominate over noise and shocks.

The theory presented in this paper could be further developed by more sophisticated derivations of the price dynamics resulting from heterogeneous actors. Alternatively, further development could address part of Kaldor's critique \cite{kaldor_irrelevance_1972}, by adding time-dependent excess demand functions. While simply adding time dependence to the potential would not capture the rich dynamics of a complex adaptive system, it would enable exploration of how markets respond in this theory to a changing economic environment.

\bibliography{why_equilibrium}

\end{document}